\documentclass[12pt]{article}
\usepackage{fullpage}
\usepackage{natbib}
\usepackage{amsmath,amssymb}
\usepackage{graphicx}
\newcommand{\startmat}[1]{\left[\begin{array}{#1}}
\newcommand{\closemat}{\end{array}\right]}

\newcommand{\bK}{\boldsymbol{K}}
\newcommand{\bW}{\boldsymbol{W}}
\newcommand{\bZ}{\boldsymbol{Z}}

\newcommand{\bA}{\boldsymbol{A}}
\newcommand{\bB}{\boldsymbol{B}}
\newcommand{\bD}{\boldsymbol{D}}
\newcommand{\bU}{\boldsymbol{U}}

\newcommand{\bs}[1]{\boldsymbol{#1}}
\newcommand{\mc}[1]{\mathcal{#1}}
\newcommand{\inner}[1]{\langle #1 \rangle}

\newcommand{\bPhi}{\bs{\Phi}}

\newcommand{\cP}{\mathbb{P}}
\begin{document}

\title{A Direct Sampler for G-Wishart Variates}
\author{Alex Lenkoski\footnote{\noindent \textit{Corresponding author address:} Alex Lenkoski, Norsk Regnesentral, P.O. Box 114 Blindern, NO-0314 Oslo, Norway
\newline{E-mail: alex@nr.no}}\\
\textit{Norwegian Computing Center}}
\maketitle
\begin{abstract}
\noindent The G-Wishart distribution is the conjugate prior for precision matrices that encode the conditional independencies of a Gaussian graphical model.  While the distribution has received considerable attention, posterior inference has proven computationally challenging, in part due to the lack of a direct sampler.  In this note, we rectify this situation. The existence of a direct sampler offers a host of new posibilities for the use of G-Wishart variates.  We discuss one such development by outlining a new transdimensional model search algorithm--which we term double reversible jump--that leverages this sampler to avoid normalizing constant calculation when comparing graphical models. We conclude with two short studies meant to investigate our algorithm's validity.
\end{abstract}
\newpage
\section{Introduction}
\indent The Gaussian graphical model (GGM) has received widespread consideration \citep[see][]{jones_et_2005} and estimators obeying graphical constraints in standard Gaussian sampling were proposed as early as \citet{dempster_1972}.  Initial incorporation of GGMs in Bayesian estimation has largely focused on decomposable graphs \citep{dawid_lauritzen_1993}, since prior distributions factorize into products of Wishart distributions.  \citet{roverato_2002} generalizes the Hyper-Inverse Wishart distribution to arbitrary graphs and, by consequence, specifies a conjugate prior for sparse precision matrices $\bK$. \citet{atay-kayis_massam_2005} further develop this prior and outline a Monte Carlo (MC) method that enables the computation of Bayes factors.  Following \citet{letac_massam_2007} and \citet{rajaratnam_et_2008}, \citet{lenkoski_dobra_2011} term this distribution the G-Wishart, and propose computational improvements to direct model comparison and model search.\\
\indent The desire to embed the G-Wishart distribution in more complicated hierarchical frameworks--particularly those involving latent Gaussianity--exposed difficulties with the MC approximation \citep[see][for discussion]{dobra_et_2011,wang_li_2012,cheng_lenkoski_2012}.  These difficulties were partly related to numerical instability \citep{wang_li_2012}, but were also methodological, as a realization of $\bK$ was needed from the current model in order to update other hierarchical parameters \citep{cheng_lenkoski_2012}.  At the time a host of Markov chain Monte Carlo (MCMC) methods had been proposed \citep{piccioni_2000,mitsakakis_et_2011,dobra_lenkoski_2011,dobra_et_2011} as well as an accept/reject sampler \citep{wang_carvalho_2010}, which \citet{dobra_et_2011} shows suffers from very low acceptance probabilities even in moderate dimensional problems.  Despite these developments no way of reliably sampling directly from a G-Wishart distribution has been proposed.\\
\indent We rectify this situation.  Our direct sampler is quite similar to the block Gibbs sampler of \citet{piccioni_2000} and involves sampling a standard Wishart variate from a full model and using the iterative proportional scaling (IPS) algorithm \citep{dempster_1972} to then place this variate in the correct space.  Our approach differs critically, however, from the block Gibbs sampler of \citet{piccioni_2000} in that sampling occurs first, and independently of previous samples, with the subsequent application of the IPS algorithm relative to a fixed target.\\
\indent The existence of a direct sampler considerably expands the usefulness of the G-Wishart distribution.  We provide one example of this, by proposing a new method of moving through the space of GGMs.  The reversible jump algorithms developed in \citet{dobra_lenkoski_2011}, and \citet{dobra_et_2011} provided a means of model averaging $\bK$ in the context of more involved Bayesian models.  As noted by \citet{wang_li_2012}, these approaches still require the use of unstable MC approximation of prior normalizing constants.  With a direct sampler, we are now able to resolve this issue by proposing a new transdimensional algorithm that combines the concept behind the exchange algorithm \citep{murray_et_2006} with reversible jump MCMC \citep{green_1995}, which we call double reversible jump.\\
\indent The article is organized as follows.  In Section~\ref{sec:gwish} we review the G-Wishart distribution, and propose the direct sampler.  Section~\ref{sec:drj} develops the new double reversible jump algorithm. In Section~\ref{sec:examples} we provide two short examples meant to confirm the validity of our new approach.  We conclude in Section~\ref{sec:conclude}.
\section{The G-Wishart Distribution}\label{sec:gwish}
\subsection{Basic Properties}
\indent Suppose that we collect data $\mc{D} = \{\bZ^{(1)}, \dots, \bZ^{(n)}\}$ such that $\bZ^{(j)}\sim \mc{N}_p(0,\bK^{-1})$ independently for $j \in \{1,\dots,n\}$, where $\bK\in\cP_p$, the space of $p\times p$ symmeteric positive definite matrices.  This sample has likelihood
$$
pr(\mc{D}|\bK) = (2\pi)^{-np/2}|\bK|^{n/2} \exp\left(-\frac{1}{2}\inner{\bK,\bU}\right),
$$
where $\inner{A,B} = tr(A'B)$ denotes the trace inner product and $\bU = \sum_{i = 1}^n \bZ^{(i)}\bZ^{(i)'}$.\\
\indent Further suppose that $G = (V,E)$ is a conditional independence graph where $V = \{1,\dots,p\}$ and $E \subset V\times V$.  As in \cite{cheng_lenkoski_2012}, we will slightly abuse notation throughout, by writing $(i,j) \in G$ to indicate that the edge $(i,j)$ is in the edge set $E$.  Associated with $G$ is a subspace $\cP_G \subset \cP_p$ such that $\bK\in \cP_G$ implies that $\bK\in\cP_{p}$ and $K_{ij} = 0$ whenever $(i,j) \not\in G$.  The G-Wishart distribution \citep{roverato_2002,atay-kayis_massam_2005} $\mc{W}_G(\delta,\bD)$ assigns probability to $\bK\in\cP_G$ as
$$
pr(\bK|\delta,\bD,G) = \frac{1}{I_G(\delta,\bD)}|\bK|^{(\delta - 2)/2}\left(-\frac{1}{2}\inner{\bK,\bD}\right)\bs{1}_{\bK\in\cP_G}.
$$
This distribution is conjugate \citep{roverato_2002} and thus
$$
pr(\bK|\delta,\bD,G,\mc{D}) = \mc{W}_G(\delta + n, \bD + \bU).
$$
\indent Let $\mc{C} = \{C_{1},\dots,C_{J}\}$ be a clique decomposition of the graph $G$.  For our purposes we assume that this decomposition is maximally complete.  We thus have that
$$
\bK_{C_j}\in\mathbb{P}_{|C_j|}
$$
for each $j \in \{1,\dots, J\}$ and $\bK\in\cP_G$.  We define the function $B_{C_j}(\cdot)$ by
$$
B_{C_j}(\bK\setminus \bK_{C_j}) = \bK_{C_{j},V\setminus C_j}\bK_{V\setminus C_j}^{-1}\bK_{V\setminus C_j, C_j}.
$$
Then given $\bK\sim\mc{W}_G(\delta,\bD)$ and any clique $C_j$ of the graph $G$, \citet{roverato_2002} proves that
\begin{equation}
\bK_{C_j}|\bK\setminus \bK_{C_j} \sim \mc{W}(\delta,\bD_{C_j},B_{C_j}(\bK\setminus \bK_{C_j})),\label{eq:full_cond}
\end{equation}
where, in general we write $\bK \sim \mc{W}(\delta,\bD,\bB)$ to denote any matrix for which $\bK - \bB \sim \mc{W}(\delta,\bD)$. Equation (\ref{eq:full_cond}) thereby gives the conditional distributions for an overlapping paritition of $E$ and proves critical to the developments below.
\subsection{Iterative Proportional Scaling and Block Gibbs Sampling}\label{sec:ips}
\indent As above, let $C_j$ be one of the cliques of $G$.  For $\bA \in \cP_{|C_j|}$ define the transformation
\begin{equation}
T_{C_j,\bA}:\cP_G\to\cP_G\label{eq:ips_function}
\end{equation}
where
$$
[T_{C_j,\bA}(\bK)]_{C_j} = \bA + B_{C_j}(\bK\setminus \bK_{C_j})
$$
while
$$
[T_{C_j,\bA}(\bK)]_{lk} = K_{lk}
$$
if either $l$ or $k$ are not in $C_j$. \citet{lenkoski_dobra_2011} use (\ref{eq:ips_function}) to determine
$$
\hat{\bK}^G = \text{argmax } |\bK|^{(\delta - 2)/2}\exp\left(-\frac{1}{2}\inner{\bK,\bD}\right)\mathbf{1}_{\bK\in\cP_G}
$$
via an algorithm known as Iterative Proportional Scaling (IPS), following the work of \citet{dempster_1972}.  The IPS algorithm works by constructing a chain $\bK^{(0)}, \bK^{(1)},\dots$ such that $\bK^{(0)} = \mathbb{I}_p$ and $\bK^{(s)}$ is determined from $\bK^{(s - 1)}$ through the update
$$
\bK^{(s)} = T_{C_J, \bD_{C_J}^{-1}} \circ \hdots \circ T_{C_1, \bD_{C_1}^{-1}}(\bK^{(s - 1)})
$$
eventually $\bK^{(s)}$ coverges to $\hat{\bK}^{G}$, see \citet{lauritzen_1996} for an in-depth discussion of the properties of the IPS algorithm.\\
\indent The IPS algorithm takes deterministic updates and therefore converges to a unique matrix.  \citet{piccioni_2000} extends the IPS idea to create an MCMC sampler for $\mc{W}_G(\delta, \bD)$.  The block Gibbs sampler of \citet{piccioni_2000} works by starting with a $\bK^{(0)}\in\cP_G$ and constructing a chain $\bK^{(1)},\bK^{(2)},\dots$ via the update
$$
\bK^{(s)} = T_{C_J, \tilde{\bK}_J} \circ \hdots \circ T_{C_1, \tilde{\bK}_1}(\bK^{(s - 1)})
$$
where $\tilde{\bK}_j$ is sampled from a $\mc{W}(\delta, \bD_{C_j})$.  We thus see that each subblock $C_j$ is being sampled from its full conditional according to (\ref{eq:full_cond}), satisfying the requirements of a Gibbs sampler.
\subsection{A Direct Sampler for G-Wishart Variates}\label{sec:direct_sampler}
We borrow ideas from Section~\ref{sec:ips} to specify a direct sampler for $\mc{W}_G(\delta,\bD)$.  First sample $\bK^{*}\sim \mc{W}(\delta, \bD)$ and determine $\bs{\Sigma} = (\bK^{*})^{-1}$.  Set $\bK^{(0)} = \mathbb{I}_p$ and construct a chain $\bK^{(1)}, \bK^{(2)},\dots$ where $\bK^{(s)}$ is updated from $\bK^{(s - 1)}$ via
\begin{equation}
\bK^{(s)} = T_{C_J, \bs{\Sigma}_{C_J}^{-1}} \circ \hdots \circ T_{C_1, \bs{\Sigma}_{C_1}^{-1}}(\bK^{(s - 1)}).\label{eq:direct_update}
\end{equation}
Eventually $\bK^{(s)}$ will converge to a matrix $\bK\in\cP_G$. We note that the key difference between our algorithm and that of \citet{piccioni_2000} is the point in which random sampling occurs.  In the block Gibbs sampler, new matrices are sampled in each step of the IPS update according to the appropriate conditional distribution.  In our framework, sampling occurs first, relative to the full model, independently of all previous samples, and the IPS is then run with a fixed target.\\
\indent The question remains what properties $\bK$ has inherited from $\bK^{*}$.  Note that by the nature of these updates, we have that 
$$
\bK_{C_j} - B_{C_j}(\bK\setminus \bK_{C_j}) = \bK^{*}_{C_j} - B_{C_j}(\bK^{*}\setminus \bK^{*}_{C_j}) = \bs{\Sigma}_{C_j}^{-1}
$$
for $j\in\{1,\dots,J\}$.  This fact is critical.  By properties of standard Wishart variates, we know that
$$
\bK^{*}_{C_j} - B_{C_j}(\bK^{*}\setminus\bK^{*}_{C_j}) \sim \mc{W}(\delta,\bD_{C_j})
$$
since this matrix has not changed, we similarly have that
$$
\bK_{C_j} - B_{C_j}(\bK\setminus\bK_{C_j}) \sim \mc{W}(\delta,\bD_{C_j})
$$
or, equivalently
$$
\bK_{C_j}|\bK\setminus \bK_{C_j} \sim \mc{W}(\delta,\bD_{C_{j}},B_{C_j}(\bK\setminus \bK_{C_j}))
$$
for all $j \in \{1, \dots, J\}$.  We note that several properties of $\bK^{*}$ are not shared by $\bK$.  For instance let $F = C_1 \cup C_2 \subset V$.  We have that
$$
\bK^{*}_F - B_{F}(\bK^{*}\setminus \bK^{*}_{F}) \sim \mc{W}(\delta, \bD_{F}),
$$
while this does not hold for $\bK$ since $K_{lk} = 0$ for any $l \in C_1$ and $k\in C_2\setminus C_1$.  Thus, while the conditional distributions of $\bK^{*}$ are not fully transferred by (\ref{eq:direct_update}), those that are relevant for $\mc{W}_G(\delta,\bD)$ variates are retained.\\
\indent The fact that $\bK\sim\mc{W}_G(\delta,\bD)$ then follows from \citet{brook_1964}.  In part, we have specified a sampler that has postive density over $\cP_G$ and the conditional distributions along a complete partition of the parameter set $E$ correspond to those of a $\mc{W}_G(\delta,\bD)$.
\subsection{Improving the Performance of the Direct Sampler}\label{sec:improved}
\indent The sampler discussed in Section~\ref{sec:direct_sampler} relied on the IPS algorithm to move from $\bK^{*}\in\cP_p$ to $\bK\in\cP_G$.  While the IPS is useful in illuminating the properties of $\bK\in\cP_G$, it is computationally burdensome.  This is for two reasons, the first of which is the requirement that the clique decomposition $\mc{C}$ be both determined and stored, an NP hard problem.  Further, the matrix $B_{C_j}(\bK\setminus \bK_{C_j})$ must be determined at each step for all $j$, an action that requires $\bK_{V\setminus C_j}$ to be solved.  If the cliques of $G$ are small, this matrix will be nearly $p\times p$.\\
\indent  \citet{hastie_et_2009} discuss an alternative algorithm to the one described in Section~\ref{sec:ips}, which can be modified to determine $\bK\in\cP_G$ from $\bK^{*}\in\cP_p$ \citep[][discuss its use in determining $\hat{\bK}^G$]{moghaddam_et_2009}.  It works in the following manner
\begin{itemize}
\item[1.] Set $\bW = \bs{\Sigma}$.
\item[2.] For $j = 1,\dots, J$
\begin{itemize}
  \item[a.] Let $N_j\subset V$ be the set of neighbors of node $j$ in $G$.  Form $\bW_{N_j}$ and $\bs{\Sigma}_{N_j,j}$ and solve
$$
\hat{\beta}^{*}_j = \bW_{N_j}^{-1}\bs{\Sigma}_{N_j,j}
$$
  \item[b.] Form $\hat{\beta}_j\in\mathbb{R}^{p-1}$ by copying the elements of $\hat{\beta}^{*}_j$ to the appropriate locations and putting zeroes in those locations not connected to $j$ in $G$.
  \item[c.] Replace $\bW_{j,-j}$ and $\bW_{-j,j}$ with $\bW_{-j,-j}\hat{\beta}_j$.
\end{itemize}
\item[3.]Repeat step 2 until convergence
\item[4.] Return $\bK= \bW^{-1}$.
\end{itemize}
\section{Double Reversible Jump}\label{sec:drj}
The direct sampler discussed in Section~\ref{sec:direct_sampler} opens the possibility for a host of new applications of the G-Wishart distribution in hierarchical Bayesian modeling.  We focus on the problem of constructing a computationally efficient algorithm for mixing over the posterior $pr(\bK,G|\mc{D})$, thereby forming a model averaged estimate of $\bK$.  We build upon the reversible jump algorithms developed in \citet{dobra_lenkoski_2011} and futher extended in \citet{dobra_et_2011}.
\subsection{Reversible Jump and Related Algorithms}
\indent Let $G$ be given and suppose that $\bK\sim\mc{W}_G(\delta + n,\bD + \bU)$.  Let $\bPhi$ be the upper triangular matrix such that $\bPhi'\bPhi = \bK$, its Cholesky decomposition. The transformation from $\bK$ to $\bPhi$ has Jacobian
$$
J(\bK\to\bPhi) = \prod_{i=1}^p\Phi_{ii}^{\nu^G_i}
$$
where $\nu_i^G = |\{j: (i,j)\in G, i < j\}|$ \citep{roverato_2002}.  Working with $\bPhi$ is useful when $\bK\in\cP_G$ since its primary restriction is that
\begin{equation}
\Phi_{ij} = -\frac{1}{\Phi_{ii}}\sum_{l = 1}^{i}\Phi_{li}\Phi_{lj}\label{eq:completion}
\end{equation}
for any $(i,j)\not\in G$. Otherwise $\Phi_{ii}\in \mathbb{R}^+$ while $\Phi_{ij}\in\mathbb{R}$ for $(i,j)\in G$.  We refer to the completion of $\bPhi$ as the action of using (\ref{eq:completion}) to augment a matrix for which only the elements of $G$ are specified.\\
\indent \citet{dobra_lenkoski_2011} use this representation to move between neighboring graphs in the context of a larger MCMC.  Suppose that $(\bK,G)$ is the current state of an MCMC chain, where $\bK\in\cP_G$ and we would like to attempt moving to $\tilde{G}$, which we assume to be equal to $G$ except for the additional edge $(l,m)$.  The algorithm of \citet{dobra_lenkoski_2011} first determines $\bPhi$ from $G$, samples $\gamma\sim \mc{N}(\Phi_{ij}, \sigma_g^2)$ and forms $\tilde{\bPhi}$ where $\tilde{\Phi}_{ij} = \Phi_{ij}$ for $i=j$ or $(i,j)\in G$, while $\Phi_{lm} = \gamma$.  $\tilde{\Phi}$ is then completed according to $\tilde{G}$.  This proposal is then accepted with probability $\min\{\alpha,1\}$ where
\begin{equation}
\alpha = \exp\left(-\frac{1}{2}\inner{\tilde{\bK} - \bK, \bD + \bU}\right)\frac{\Phi_{ll}\sqrt{2\pi}\sigma_g}{\exp(-\frac{\gamma^2}{2\sigma_g^2})}\frac{I_{\tilde{G}}(\delta,\bD)}{I_{G}(\delta,\bD)}\label{eq:dl_rj}.
\end{equation}
Subsequent to this move, the matrix $\bK$ has typically been updated according to the accepted graph using MCMC methods, for instance the block Gibbs sampler.\\
\indent Several embellishments of the algorithm of \citet{dobra_lenkoski_2011} have been developed, including asymmetric model moves in the graph space and permuting the elements of $\bK$ to increase acceptance \citep{dobra_et_2011}, noting that a conditional Bayes factor can be derived to obviate the need for reversible jump when comparing neighboring graphs \citep{wang_li_2012} and using notions of sparse Cholesky decompositions and node reorderings to reduce the time spent computing $\bPhi$ \citep{cheng_lenkoski_2012}.\\
\indent Each of these developments has proven to yield some improvement in performance in certain situations.  However, the most important technical problem with (\ref{eq:dl_rj}) revolves around the calculation of the normalizing constants $I_G$ and $I_{\tilde{G}}$.  These factors require MC approximation \citep{atay-kayis_massam_2005}, which \citet{wang_li_2012} rather convincingly show fails in high dimensions.\\
\indent \citet{wang_li_2012} propose an alternative approach, which borrows ideas from the exchange algorithm \citep{murray_et_2006} and the double Metropolis-Hastings algorithm \citep{liang_2010} to approximate this ratio. Unfortunately, the double Metropolis-Hastings algorithm  is not exact, though the approximation used by \citet{wang_li_2012} appears to work well in practice for neighboring graphs.  We note that the approach of \citet{wang_li_2012} is not feasible if the graphs are not neighbors.
\subsection{The Double Reversible Jump Algorithm}\label{sec:drj_first}
\indent The exchange algorithm \citep{murray_et_2006} has proven a useful tool for general MCMC when working with models where the likelihood has an intractable normalizing constant.  \citet{wang_li_2012} discuss how to use the concept behind the exchange algorithm to aid model comparison, where prior distributions--like those of the G-Wishart--have a similar unknown normalizing constant that varies according to the model.  Unfortuntately, without a direct sampler \citet{wang_li_2012} relied on the block Gibbs sampler to propose a version of the double Metropolis-Hastings algorithm \citep{liang_2010}.  This approach should only be considered approximate, whereas the original exchange algorithm avoids normalizing constant calculations and still yields correct MCMC transition probabilities.\\
\indent With the existence of a direct sampler for $\mc{W}_G$ variates, however, we may use a modification of the exchange algorithm to avoid the normalizing constants in (\ref{eq:dl_rj}).  We call this new approach double reversible jump.\\
\indent Suppose that $G$ is the graph in the current state of an MCMC procedure and propose a new graph $\tilde{G}$. At the moment, assume that $\tilde{G}$ is a neighbor of $G$ with the additional edge $(l,m)\in\tilde{G}$.  We discuss the relaxation of this assumption in Section~\ref{sec:conclude}.  The double reversible jump algorithm then proceeds by
\begin{itemize}
\item[1.] Sample $\bK \sim \mc{W}_G(\delta+n,\bD+ \bU)$ and form $\bPhi$, its Cholesky decomposition.  Let $\vartheta = \Phi_{lm}$.
\item[2.] Sample $\tilde{\bK}^0 \sim\mc{W}_{\tilde{G}}(\delta,\bD)$ and form $\tilde{\bPhi}^0$.  Let
$$
\tilde{\vartheta} = -\frac{1}{\tilde{\Phi}^0_{ll}}\sum_{r = 1}^{l}\tilde{\Phi}^0_{rl}\tilde{\Phi}^0_{rm}
$$
\item[3.] Sample $\gamma\sim\mc{N}(\vartheta,\sigma_g^2)$ and set $\tilde{\gamma} = \tilde{\Phi}^0_{lm} - \tilde{\vartheta}$
\item[4.] Form $\tilde{\bPhi}$ where $\tilde{\Phi}_{ij} = \Phi_{ij}$ for $(i,j)\in G$ or $i=j$ and set $\tilde{\Phi}_{lm} = \gamma$.  Complete $\tilde{\bPhi}$ according to $\tilde{G}$ and set $\tilde{\bK} = \tilde{\bPhi}'\tilde{\bPhi}$
\item[5.] Form $\bPhi^0$ where $\Phi^0_{ij} = \tilde{\Phi}_{ij}^0$ for $(i,j)\in G$ and $\Phi_{lm}^0 = \tilde{\vartheta}$.  Complete $\bPhi^0$ according to $G$ and set $\bK^0 = (\bPhi^0)'\bPhi^0$.
\item[6.] Accept the move from $G$ to $\tilde{G}$ with probability $\min\{1,\alpha\}$ where
$$
\alpha = \frac{\exp\left(-\frac{1}{2}\inner{\tilde{\bK} - \bK, \bD + \bU}\right)}{\exp\left(-\frac{1}{2}\inner{\tilde{\bK}^0 - \bK^0,\bD}\right)}\frac{\Phi_{ll}}{\Phi_{ll}^0}\exp\left(-\frac{(\gamma-\vartheta)^2 - (\tilde{\gamma}-\tilde{\vartheta})^2}{2\sigma^2_g}\right)
$$
\end{itemize}
We see that the double reversible jump algorithm considers switching between
$$
(\bK,G,\tilde{\bK}^{0}, \tilde{G})
$$
to the alternative
$$
(\tilde{\bK}, \tilde{G}, \bK^0, G)
$$
by performing two reversible jump moves, one that moves between $(\bK, G)$ to $(\tilde{\bK}, \tilde{G})$ according to the posterior parameters $\delta + n$ and $\bD + \bU$ and the other between $(\tilde{\bK}^0, \tilde{G})$ to $(\bK^0, G)$ according to the prior parameters $\delta$, $\bD$.  By doing so, the prior normalizing constants in (\ref{eq:dl_rj}) cancel, making double reversible jump the transdimensional equivalent to the exchange algorithm of \citet{murray_et_2006}.
\section{Examples}\label{sec:examples}
\subsection{Sampling from a fixed, low-dimensional model}
\indent We begin with a simple sanity check to ensure that the direct sampler of Section~\ref{sec:direct_sampler} returns identical results as the block Gibbs sampler when both are run for an exceedingly long time.  We set $p=4$ and $G = C_4$ the four cycle where edges $(1,4)$ and $(2,3)$ are missing.  We then consider sampling from $\mc{W}_{C_4}(\delta,\bD)$ where we set $\delta = 103$ and 
$$
\bD = \startmat{cccc}136.431 & -10.15 & 8.027 & 2.508\\
-10.15 & 93.417 & -2.122 & -16.162\\
8.027 & -2.122 & 116.652 & 11.62\\
2.508 & -16.162 & 11.62 & 120.203\closemat
$$
which was randomly generated to resemble the posterior distribution after observing $100$ samples drawn from a $\mc{N}_4(0,\mathbb{I}_4)$.  We then ran the block Gibbs sampler as well as the direct sampler for $10$ million iterations each, with an additional one million iterations for the block Gibbs sampler as burn-in.  The expection of $\bK$ taken over the samples from the block Gibbs sampler was
$$
\startmat{cccc}0.7788 & 0.0827 & -0.0516 & 0\\
0.0827 & 1.1594 & 0 & 0.1528\\
-0.0516 & 0 & 0.9122 & -0.0864\\
0 & 0.1528 & -0.0864 & 0.9025\closemat.
$$
the expectation of $\bK$ taken over samples from the proposed direct sampler was
$$
\startmat{cccc}0.7788 & 0.0826 & -0.0516 & 0\\
0.0826 & 1.1593 & 0 & 0.1527\\
-0.0516 & 0 & 0.9122 & -0.0863\\
0 & 0.1527 & -0.0863 & 0.9024\closemat
$$
As shown above, the expectations of the two samplers appear identical.  Further, we note that all other comparisons we could consider--for instance element-wise variance, quantiles of determinants, medians--likewise returned identical results.  Different fixed graphs and choices of $\delta$ or $\bD$ do not affect the results.\\
\indent Since 10 million samples of the block Gibbs sampler after a one million sample burn-in should be expected to characterize a $\mc{W}_{C_4}(\delta,\bD)$ distribution, this brief study appears to confirm that our proposed sample is indeed a direct sampler for G-Wishart variates.
\subsection{Fisher's Iris Data}
\indent Both \citet{roverato_2002} and \citet{atay-kayis_massam_2005} use Fisher's \emph{Iris Virginica} dataset to confirm their approximations of $I_G$.  These data consist of four measurements, Sepal Length (SL), Sepal Width (SW), Petal Length (PL) and Petal Width (PW), taken on 50 iris plants.  We use these data to compare the double reversible jump algorithm to an exhaustive scoring of all models using the Monte Carlo approximation in \citet{atay-kayis_massam_2005}.  We set $\delta = 3$, $\bD = \mathbb{I}_p$ and $\sigma^2_g = 1$.\\
\indent For all $64$ models in the graph space, we determine the model probability by running the MC approximation of \citet{atay-kayis_massam_2005} for one-million iterations.  Further, we run the double reversible jump algorithm for five-million iterations and discard the first 100,000 iterations as burn-in.  This takes approximately ten minutes on a 2.8 gHz desktop running Linux and should be recognized as an extremely long chain.  Table~\ref{tab:iris} shows the pairwise edge probabilities returned from the two methods.  As shown in the table, the estimated edge probabilities using the two approaches agree.  We note that if the double reversible jump chain is run for less time, say 50,000 iterations (which takes approximately 6 seconds), results are nearly, but not perfectly, identical.  Model moves are accepted in 23.2\% of attempts, alternative choices of $\sigma^2_g$ appear to have marginal effect on this acceptance level.   See Section~\ref{sec:conclude} for a discussion of ways to improve mixing through more involved schemes.\\
\indent These results indicate that the double reversible jump, coupled with the direct sampler, enables model averaged estimates of $\bK$ to be formed without needing the unstable MC approximation of \citet{atay-kayis_massam_2005}.
\section{Conclusions}\label{sec:conclude}
\indent We have proposed a direct sampler for G-Wishart variates, which promises to dramatically improve the usefulness of this distribution.  In this note we have focused on using this sampler to develop a trandimensional MCMC algorithm that has no normalizing constant evaluations.  While this is a promising first step, there are considerable additional avenues for development.\\
\indent While the direct sampler performs well, in our mind the entire process is still too slow.  In high dimensions, the majority of computing time is spent moving from $\bK^{*} \in \cP_p$ to $\bK\in\cP_G$.  While the development in section~\ref{sec:improved} is considerably faster (and dramatically more stable) than the use of the IPS algorithm, we feel that there must be potiential for further improvements.  Connecting with the rapid development of procedures for forming glasso \citep{friedman_et_2008} estimators would be fruitful in improving the efficiency of the sampler in high dimensions, since this action can be phrased as a constrained optimization problem.\\
\indent \citet{rodriguez_et_2011} consider embedding the G-Wishart distribution inside Dirichlet processses and related structures from nonparameteric Bayesian methods.  However, decomposable graphs were used, since a direct sampler was unavailable for nondecomposable models and is critical in the posterior sampling of nonparameteric models.  It is now possible to  consider the use of general graphical models in Bayesian nonparametric approaches.\\
\indent Our development of the double reversible jump algorithm was partially to show how the direct sampler could be used to avoid prior normalizing constant evaluations when comparing models.  A host of embellishments could be made.  When comparing neighboring graphs, for instance, conditional Bayes factors could be computed as in \citet{wang_li_2012} or \citet{cheng_lenkoski_2012}.  In our mind, a more promising avenue for development would be to construct a procedure for global moves in the graph space.  In order to work properly, we feel that global moves must be coupled with better proposals in the double reversible jump scheme.  Relating these proposals to some of the guidelines in \citet{rue_held_2005} could prove useful in this regard.  The ability to make large, focused moves in the graph space will be critical to extending the G-Wishart distribution to truly high dimensional problems.
\bibliographystyle{apalike}
\bibliography{Lenkoski}
\section*{Acknowledgement} Alex Lenkoski's work is funded by Statistics for Innovation $(sfi)^2$, in Oslo.  The author gratefully acknowledges helpful discussion with Arnoldo Frigessi, Adrian Dobra, and the participants of a mini-workshop at NTNU in Trondheim held on March 13th, 2013.
\newpage
\begin{table}
\caption{Pairwise edge probabilities from Monte Carlo (lower triangle) and double reversible jump (upper triangle) in the iris dataset}\label{tab:iris}
\begin{center}
\begin{tabular}{lcccc}
\hline\hline
 & SL & SW & PL & PW\\
\hline
SL & 1 & 0.821 & 1 & 0.405\\
SW & 0.821 & 1 & 0.501 & 0.987\\
PL & 1 & 0.501 & 1 & 0.532\\
PW & 0.406 & 0.987 & 0.532 & 1\\
\hline\hline
\end{tabular}
\end{center}
\end{table}

\end{document}